\newcommand{\diracslash}[1]{#1\llap{/\kern2pt}}

\newcommand{\be}{\begin{equation}}
\newcommand{\ee}{\end{equation}}
\newcommand{\bea}{\begin{eqnarray}}
\newcommand{\eea}{\end{eqnarray}}
\newcommand{\ba}[1]{\begin{array}{#1}}
\newcommand{\ea}{\end{array}}

\newcommand{\bt}{\begin{tabular}}
\newcommand{\et}{\end{tabular}}

\newcommand{\beas}{\begin{eqnarray*}}
\newcommand{\eeas}{\end{eqnarray*}}

\documentclass[preprint,prd,aps,floats,nofootinbib,showpacs,floatfix]{revtex4}
\usepackage{graphicx}
\begin{document}

\title{$\rho $ - Meson spectral function in hot \linebreak asymmetric nuclear matter}
\author{P.C.Raje Bhageerathi}
\email{pc_raje@yahoo.com}
\affiliation{Department of Physics,Providence Women's College, Calicut-673002,
Kerala, India}

\begin{abstract}

The $\rho$-meson spectral function in hot and dense asymmetric 
nuclear matter (number densities of proton and neutron different) is evaluated in 
an effective chiral SU(3) model within the mean field 
approximation (MFA). The dependence of the vector meson masses on density 
and tempertaure, their variation with the asymmetry parameter, the form of 
the spectral function when the $p\longleftrightarrow n $ symmetry is broken  are studied.
One can observe a clear splitting of the longitudinal and transverse
modes among the $\rho$ isospin multiplets. The role of a running tensor 
coupling on the spectral function is also studied.

\end{abstract}
\maketitle

\def\bfm#1{\mbox{\boldmath $#1$}}

\section{Introduction}
The behaviour of vector mesons in nuclear matter under extreme conditions 
has attracted a lot of attention during the recent years, both theoretically and experimentally. 
The properties of hadrons at high temperatures and densities are quite different from 
the properties of hadrons in vacuum. The medium modification of hadrons have direct 
consequences on the experimental observables from the strongly interacting matter 
produced in heavy ion collision experiments. The reduction of the vector meson masses 
in hot and/or dense nuclear matter could be regarded as a possible signal of the restoration 
of chiral symmetry \cite{Brown, Hat1}. The study of the $ \rho $ meson in the medium is 
particularly interesting, because of the fact that it can decay directly to lepton pairs \cite{Li, Chanf}.

The in-medium properties of the vector mesons have been extensively studied within
the framework of different models \cite{Wal1, Ser1, Ser2}. Since different formalisms
lead to different results, there exists a lot of controversy regarding these issues.
The Brown and Rho \cite{Brown} scaling suggests a dropping of the vector meson masses
in the medium. Gale and Kapusta \cite{Gale} have earlier analyzed the modification of
the $ \rho $ meson self energy at finite temperature in an isospin-symmetric pion 
medium. They have observed the medium corrections to be modest even upto a temperature
of about $ 0.15  $ GeV. Chin \cite{Chin} in his pioneering work predicted an increase in the 
mass of the $ \omega $ meson in the dense nuclear matter within the mean field approximation
(MFA), where the contribution from the Dirac sea ($ n-\bar{n} $ pair production) is not 
considered. The mass of the $ \omega $ meson has been found to decrease in dense nuclear
matter by Jean et al. \cite{Jean} by taking into account of both the Fermi and the Dirac
sea effects. Within the framework of QHD-I and chiral SU(3)models, it is reported that the 
peak of the spectral function shifts towards lower invariant mass regions in RHA, while
it is centered around the nominal $ \rho $-pole in MFA \cite{PCR} in isospin symmetric medium.     
The work by Gale and Kapusta has been extended to an isospin-asymmetric pion
medium with large values of the charge chemical potential $ \mu $ \cite{Gul}. They have
predicted an asymmetry in the emission rate of the dileptons. The effects of the asymmetry parameter
on the masses of the $ \rho^{0} $, $ \rho^{+} $ and the $ \rho^{-} $- mesons is
studied in \cite{Mazumder}, where it has been shown that the $ \rho^{+} $ -meson 
takes slightly higher and the $ \rho^{-} $-meson takes slightly lower masses
compared to the $ \rho^{0} $, with increasing values of the nucleon density.      
The propagation of the light mesons $\sigma, \omega, \rho $, and a0(980) in dense hadronic matter is investigated
in an effective relativistic hadronic model inspired in the derivative scalar coupling model (DCM) \cite{Agu}.
   
In this paper, we attempt to study the properties of the $ \rho $ meson by considering its 
propagation in hot and dense asymmetric nuclear matter focussing on the role of 
$p\longleftrightarrow n $ symmetry breaking using an effective Lagrangian in chiral SU(3) model
within the mean field approximation (MFA). Chiral SU(3) models have been widely used in the study of nuclear matter,
finite nuclei, and hyperonic matter. The properties of the vector mesons in nuclear medium,
and the energies of the kaons (antikaons) in the asymmetric nuclear matter were also 
studied using this model \cite{AM}. In the present work we have not considered the 
nucleon-antinucleon ($ n-\bar{n} $ ) excitation effect. There is an enhancement of effective mass
for $ \rho^{+,0} $, whereas a reduction is seen for $ \rho^{-} $ at the maximum value
of the asymmetry (neutron matter). The fact that the neutron and proton densities are different, 
brings a modification of the propagation of $ \rho^{\pm,0} $ mesons in hot and dense asymmetric nuclear matter.
The remarkable splitting of the spectral functions of the $ \rho $ meson triplets, when the discrete 
isospin symmetry is broken is revealed.

We organize the paper as follows: Section 2 gives a brief
description of the chiral SU(3) model used in the 
present investigation. Section 3 discusses
the effects of finite temperature, asymmetry and density on the spectral  
function of the $\rho$ meson. Section 4 contains the results
and discussion and in section 5, we summarize our main findings of the present investigation.

\section{The hadronic chiral SU(3)$ \times $ SU(3) model }
The effective hadronic chiral Lagrangian density used in the present work is given as
\begin{equation}
L=L_{kin}+\sum_{W=X,Y,V,A,u} L_{BW} +L_{vec}+L_{0}+L_{SB}
\end{equation}
Eq. (2.1) corresponds to a relativistic model of baryons and mesons adopting a 
nonlinear realization of chiral symmetry \cite{Wein,Cole,Call,Bar} and broken scale 
invariance \cite{paper3, Zsch1, Zsch2} as a description of the hadronic matter. 
The model was used successfully to describe nuclear matter, finite nuclei, 
hypernuclei and neutron stars. The Lagrangian contains the baryon octet, the spin-0
and spin-1 meson multiplets as the elementary degrees of freedom. Here, $L_{kin}$ is 
kinetic energy term, $L_{BW}$ is the baryon-meson interaction term in which the 
baryons-spin-0 meson interaction term generates the baryon masses. $L_{vec}$  
describes the dynamical mass generation of the vector mesons via couplings to the 
scalar mesons and contains additionally quartic self-interactions of the vector fields. 
$L_{0}$ contains the meson-meson interaction terms inducing the spontaneous breaking 
of chiral symmetry as well as a scale invariance breaking logarithmic potential. $L_{SB}$ 
describes the explicit chiral symmetry breaking. 

Baryon-scalar meson interactions generate the baryon masses through the 
coupling of the baryons to the non-strange $ \sigma $, the strange $ \zeta $ scalar  
mesons and also to scalar-isovector meson $ \delta $. The $ \delta $ meson is responsible 
for splitting of proton and neutron effective masses. 
The parameters $ g_{1}^{S} $, $ g_{8}^{S} $ and $\alpha_{S}$ 
are adjusted to fix the baryon masses to their experimentally measured 
vacuum values. It should be emphasized that the nucleon mass also depends 
on the \textit{strange condensate} $ \zeta $. For the special case of 
ideal mixing $ (\alpha_{S}=1 $ and $ g_{1}^{S}=\sqrt{6}g_{8}^{S})$ 
the nucleon mass depends only on the non-strange quark condensate.

In analogy to the baryon-scalar meson coupling there exist two independent 
baryon-vector meson interaction terms corresponding to the F-type 
(antisymmetric) and D-type (symmetric) couplings. Here we will use 
the antisymmetric coupling, because the universality principle 
\cite{saku69} and the vector meson dominance model suggest that 
the symmetric coupling should be small.  Additionally, we choose the parameters \cite{Wein,Cole,Bar} 
so as to decouple the strange vector field $ \phi_{\mu}\sim \bar{s}\gamma_{\mu}s $
from the nucleon, corresponding to an ideal mixing between
$\omega $ and the $\phi $ -mesons. A small deviation of the mixing angle from ideal mixing has not
been taken into account in the present investigation.

The concept of broken scale invariance leading to the trace anomaly in (massless) QCD,
$ \theta_{\mu}^{\mu}=(\beta_{QCD}/2g)\left\langle G^{a}_{\mu\nu}
G^{a,\mu\nu}\right\rangle  $, where $ G^{a}_{\mu\nu} $ is the gluon field 
strength tensor of QCD, is simulated in the
effective Lagrangian at tree level \cite{paper3, Zsch1, Zsch2} through the introduction 
of the scale breaking terms
\begin{equation}
L_{scalebreak} = -\frac{1}{4}\chi^{4}\ln \frac{\chi^{4}}{\chi_{0}^{4}}+\frac{d}{3}\chi^{4}\ln \Big(\Big(\frac{I_{3}}{det\left\langle X\right\rangle _{0}}\Big)\Big( \frac{\chi}{\chi_{0}}\Big)^{3}\Big)
\end{equation}
where $ \textit{I}_{3}= \textit{det(X)} $ with $ X $ as the multiplet for the scalar mesons.
The effect of these logarithmic terms is to break the scale
invariance, which leads to the trace of the energy momentum tensor as 
\begin{equation}
\theta_{\mu}^{\mu}= (1-d)\chi^{4} .
\end{equation}
Hence the scalar gluon condensate of QCD ( $\left\langle G^{a}_{\mu\nu}
G^{a,\mu\nu}\right\rangle  $) 
is simulated by a scalar dilaton
field in the present hadronic model.

The hadronic properties are studied in the hot and dense asymmetric medium within the mean field approximation,
where all the meson fields are treated as classical fields. Also in this approximation, only the scalar and the vector fields contribute to the baryon-meson interaction, $ L_{BW}$,  
since for all other mesons, the expectation values are zero. The effects of the isospin asymmetry is introduced through  
the scalar-isovector $ \delta $ field and the isospin asymmetry parameter $ \eta $ is defined as
$ \eta = \frac{1}{2} (\rho_{N}-\rho_{P})/\rho_{B}$, where $ \rho_{N} $ and $ \rho_{P} $ are the number densities of the 
neutron and the proton and   $ \rho_{B}=\rho_{N}+\rho_{P} $, is the total baryon density. 

The Lagrangian density in the MFA has the following terms: 
\begin{eqnarray}
L_{BX}+ L_{BV}  = & &-\sum_{i} \bar{\psi}_{i}\left[ g_{i\omega}\gamma_{0}\omega +g_{i\rho}\gamma_{0}\rho + g_{i\phi}\gamma_{0}\phi+ M_{i}^{\ast}\right] \psi_{i} , \nonumber\\
L_{vec} = & & \frac{1}{2}\frac{\chi^{2}}{\chi_{0}^{2}}(m_{\omega}^{2}\omega^{2} +m_{\rho}^{2}\rho^{2}+m_{\phi}^{2}\phi^{2})+ g_{4}^{4}(\omega^{4}+6\omega^{2}\rho^{2}+\rho^{4}+2\phi^{4}) ,\nonumber\\
L_{0} = & & -\frac{1}{2}k_{0}\chi^{2}\left( \sigma^{2}+\zeta^{2}+\delta^{2}\right)+ k_{1} \left( \sigma^{2}+\zeta^{2}+\delta^{2}\right)^{2} + k_{2}
\left( \frac{\sigma^{4}}{2}+\frac{\delta^{4}}{2}+\zeta^{4}+3\sigma^{2}\delta^{2}\right)\nonumber\\
& & + k_{3}\chi(\sigma^{2}-\delta^{2})\zeta -  k_{4}\chi^{4} -  \frac{1}{4}\chi^{4}\ln \frac{\chi^{4}}{\chi_{0}^{4}}
+ \frac{d}{3}\chi^{4}\ln \Big(\frac{(\sigma^{2}-\delta^{2})\zeta}{\sigma_{0}^{2}\zeta_{0}}\Big(\frac{\chi}{\chi_{0}}\Big)^{3}\Big) , \nonumber\\
L_{SB} = & & -\left( \frac{\chi}{\chi_{0}}\right) ^{2}\left[ m_{\pi}^{2}f_{\pi}\sigma + \left( \sqrt{2}m_{K}^{^{2}} f_{K} - \frac{1}{\sqrt{2}}m_{\pi}^{2}f_{\pi}\right) \zeta\right]  ,
\end{eqnarray}
where $ M_{i}^{\ast} $ = $ -g_{\sigma i} \sigma -g_{\zeta i}\zeta -g_{\delta i}\delta $  is the effective mass of the baryon of species i. Here $ k_{0} $, $ k_{1} $, $ k_{2} $, $ k_{3} $, $ k_{4} $, $ \delta $ 
and $ g_{4} $ are parameters corresponding to the Mean-field and the Hartree approximations 
in the chiral SU(3) model. The thermodynamical potential of the grand canonical ensemble  $ \Omega $  per unit volume V at a given chemical potential 
$ \mu_{i} $ and temperature T can be written as
\begin{eqnarray}
\frac{\Omega}{V} = & & - L_{vec} - L_{0} - L_{SB} - \nu_{vac} - T\frac{\gamma_{i}}{(2\pi)^{3}}\int d^{3}p \Big[\ln \Big(1+ e^{-\frac{1}{T}[E_{i}^{*}(p)-\mu_{i}^{*}]}\Big)\Big]. 
\end{eqnarray}
Here $ \nu_{vac} $  is the vacuum energy (the potential at $ \rho $ = 0) that has 
been subtracted inorder to obtain a vanishing vacuum energy \cite{paper3}, $ \gamma_{i} $ are the spin-isospin degeneracy factor, and $ \gamma_{i} $ = 2
for asymmetric nuclear matter. 
where, $ E_{i}^{*} = \sqrt{p^{2}+M_{i}^{*2}} $ and $ \mu_{i}^{*} = \mu_{i}-g_{i\omega}\omega-g_{i\rho}\rho-g_{i\phi}\phi$ are the single particle energy and the effective chemical potential for the nucleon of species i. The mesonic field equations are determined by minimizing the thermodynamic potential.
We shall use the frozen glueball approximation $ \left( \chi = \chi_{0}\right) , $ since the dilaton field which simulates the gluon condensate changes very little in the medium. We then have 
coupled equations only for the fields $ \sigma,\zeta $, $ \delta $  and  $ \omega $ 
as given by
\begin{eqnarray}
\frac{\partial\left( \Omega/V\right) }{\partial\sigma} = & & k_{0}\chi^{2}\sigma - 4k_{1}\left( \sigma^{2}+\zeta^{2}+\delta^{2}\right) \sigma - 2k_{2}(\sigma^{3}+3\sigma\delta^{2}) - 2k_{3}\chi\sigma\zeta - \frac{d}{3}\chi^{4}\Big(\frac{2\sigma}{\sigma^{2}-\delta^{2}}\Big) \nonumber\\
& & + m_{\pi}^{2}f_{\pi} +\sum_{i} \frac{\partial M_{i}^{\ast}}{\partial\sigma}\rho_{i}^{S} = 0,
\end{eqnarray}
\begin{eqnarray}
\frac{\partial\left(\Omega/V \right) }{\partial\zeta} = & & k_{0}\chi^{2}\zeta - 4k_{1}\left( \sigma^{2}+\zeta^{2}+\delta^{2}\right) \zeta - 4k_{2}\zeta^{3} - k_{3}\chi(\sigma^{2}-\delta^{2}) - \frac{d\chi^{4}}{3\zeta}+ \frac{\partial M_{i}^{\ast}}{\partial\zeta}\rho_{i}^{S}  \nonumber\\
& & + \left[ \sqrt{2}m_{K}^{2}f_{K}- \frac{1}{\sqrt{2}}m_{\pi}^{2}f_{\pi}\right]  = 0, 
\end{eqnarray}
\begin{eqnarray}
\frac{\partial\left( \Omega/V\right) }{\partial\delta} = & & k_{0}\chi^{2}\delta - 4k_{1}\left( \sigma^{2}+\zeta^{2}+\delta^{2}\right) \delta -2 k_{2}(\delta^{3}+3\delta\sigma^{2}) + k_{3}\chi\delta\zeta - \frac{d}{3}
\chi^{4}\Big(\frac{2\delta}{\sigma^{2}-\delta^{2}}\Big) \nonumber\\
& & +\sum_{i} \frac{\partial M_{i}^{\ast}}{\partial\delta}\rho_{i}^{S} = 0,
\end{eqnarray}
\begin{equation}
\frac{\partial\left( \Omega/V \right) }{\partial\omega} = - m_{\omega}^{2}\omega - 4g_{4}^{4}\omega^{3} + g_{N\omega}\rho_{i} = 0,
\end{equation}

which have to be solved self-consistently to obtain the values of $ \sigma $, $ \zeta $, $ \delta $ and $ \omega $ .      
Here $ \rho_{i}^{S} $  and  $ \rho_{i } $ are the scalar and vector densities
for the nuclear matter at finite temperature, T given as
\begin{equation}
\rho_{i}^{S} =  \gamma_{i}\int \frac{d^{3}p}{(2\pi)^{3}}\frac{M_{i}^{*}}{E_{i}^{*}}(n_{i}(p)+\bar{n}_{i}(p)), \;\;\;\;
\rho_{i} =  \gamma_{i}\int \frac{d^{3}p}{(2\pi)^{3}}(n_{i}(p)-\bar{n}_{i}(p))
\end{equation}
The $ n_{i}  $ and $ \bar{n}_{i} $ are the thermal 
distribution functions for the neutron (proton) and the antineutron (antiproton) respectively. 
\begin{equation}
n_{i}(p) = \frac{1}{e^{(E_{i}^{*}-\mu_{i}^{*})/T}+1},   \qquad 
\bar{n}_{i}(p) = \frac{1}{e^{(E_{i}^{*}+\mu_{i}^{*})/T}+1}  .  
\end{equation}

\section{Rho meson spectral function in asymmetric nuclear matter}

The hadronic matter produced in ultra-relativistic heavy-ion collisions at very high 
densities and temperatures is studied using the finite temperature field theory \cite{Das, Bel}. 
In the present calculation thermal effects enter through thermal neutron and proton 
loops. In Minkowski space, the self-energy of the $ \rho $
vector meson can be expressed as \cite{Gale,Kap}
\begin{eqnarray}
\Pi^{\mu\nu}\left( k\right) & = & \Pi_{L}\left( k\right)P_{L}^{\mu\nu} + \Pi_{T}\left( k\right)P_{T}^{\mu\nu},
\end{eqnarray}
where $ k^{2} $ = $ k_{0}^{2}-|\vec{k}^{2}| $. The $ P_{L}^{\mu\nu} $ and $ P_{T}^{\mu\nu}  $
are the longitudinal and transverse projection tensors defined as
\begin{eqnarray}
P_{T}^{00}=P_{T}^{0i}=P_{T}^{i0}=0 ,\: P_{T}^{ij}=\delta^{ij}-\frac{k_{i}k_{j}}{|\vec{k}^{2}|} ,\: 
& P_{L}^{\mu\nu}=\frac{k^{\mu}k^{\nu}}{k^{2}}-g^{\mu\nu}-P_{T}^{\mu\nu}.
\end{eqnarray}
$ \Pi_{L} $ and $ \Pi_{T} $ are related to the components of the self-energy by \cite{Song}
\begin{eqnarray}
\Pi_{L}\left( k\right) = \frac{k^{2}}{|\vec{k}^{2}|}\Pi^{00}\left( k\right) ,  \qquad 
\Pi_{T}\left( k\right) = -\frac{1}{2}\left( \Pi_{\mu}^{\mu}+\Pi_{L}\left( k\right) \right) .
\end{eqnarray}

The imaginary part of the retarded propagator is referred to as the 
spectral function. The study of the $ \rho $ meson spectral function is attributed to calculating the
in-medium self-energy of the $ \rho $ meson.

\subsection{$ \rho NN $ interaction}

The contribution of nucleon excitations through nucleon-loop
to $ \rho $ self-energy is analyzed in terms of the effective
Lagrangian density \cite{Hat2}
\begin{equation}
L_{\rho NN} =  g_{\rho NN}\left( \bar{\Psi}\gamma_{\mu}\tau^{a}\Psi V_{a}^{\mu} - \frac{\kappa_{\rho}}{2M_{i}}\bar{\Psi}\sigma_{\mu\nu}\tau^{a}\Psi\partial^{\nu}V_{a}^{\mu}\right) ,
\end{equation}
where $ V_{a}^{\mu} $ is the $ \rho $ meson field and $ \Psi $ is the nucleon 
field. The second order polarization function or the self energy can be written as
\begin{equation}
\Pi_{\mu\nu}^{\rho NN} = -ig_{\rho NN}^{2}\int\frac{d^{4}p}{(2\pi)^{4}}Tr\left[ \Gamma_{\mu}\left( k\right) G(p+k) \Gamma_{\nu}\left( -k\right) G(p)\right] , 
\end{equation}
where
\begin{equation}
\Gamma^{\mu}\left( k\right) = \gamma^{\mu}+\frac{i\kappa_{\rho}}{2M_{i}}\sigma_{\mu\nu}k^{\nu},
\end{equation}
with $ \sigma_{\mu\nu}=\frac{i}{2}[\gamma_{\mu},\gamma_{\nu}] $,
$ M_{i} $ and $ M_{i}^{*} $ are the neutron (proton) masses in vacuum and in the hot hadronic medium
respectively. $G(p)$ is the nucleon propagator in matter which can be written as
\begin{equation}
G(p)=(p+M_{i}^{*})\left\lbrace \frac{1}{p^{2}-M_{i}^{*2}}+i\pi\delta(p^{2}-M_{i}^{*2})\left[\theta(p_{0})n_{i}
(p)+\theta(-p_{0})\bar{n_{i}}(p)\right]\right\rbrace .    
\end{equation}

The polarization tensor $ \Pi_{\mu\nu} ^{\rho NN}\left( k\right) $
can be separated into two parts,
\begin{equation}
\Pi_{\mu\nu} ^{\rho NN}\left( k\right) = -\left(g_{\mu\nu}- \frac{k_{\mu}k_{\nu}}{k^{2}}\right)\Pi_{F}^{\rho NN}\left( k\right)\
 + \Pi_{D,\mu\nu}^{\rho NN}\left( k\right) ,
\end{equation}
corresponding to the vacuum and the matter contributions.
Using dimensional regularization and taking a phenomenological
subtraction procedure \cite{Hat2}, the vacuum part (T = 0) is,
\begin{equation}
\Pi_{iF, L\left( T\right) }^{\rho NN}\left( k\right) = k^{2}\left( \frac{g_{\rho NN}}{\pi}\right) ^{2}\left( I_{1}+\frac{\kappa_{\rho}M_{i}^{\ast}}{2M_{i}}I_{2} + \left( \frac{\kappa_{\rho}}{2M_{i}}\right) ^{2}\frac{k^{2}I_{1}+M_{i}^{\ast 2}I_{2}}{2}\right) ,
\end{equation}
where
\begin{equation}
I_{1}=\int_{0}^{1} dx\,x\,\left( 1-x\right) \ln C  ,  I_{2}=\int_{0}^{1} dx \ln C ,   C=\frac{M_{i}^{\ast 2}-x\left( 1-x\right) k^{2}}{M_{i}^{2}-x\left( 1-x\right) k^{2}}.
\end{equation}

The polarization tensor for $ \rho^{(0,\pm)} $ will be same for symmetric nuclear matter and the condition for current conservation ($ k^{\mu}\Pi_{\mu\nu}= \Pi_{\mu\nu}k^{\nu}=0 $) is automatically fulfilled. Even when the nuclear matter is asymmetric, the current is conserved for $ \rho^{0} $ as it involves only the $ p-p $ and $ n-n $ loops. i.e., $ \rho^{0} $ is blind to isospin asymmetry. But for the $ \rho^{\pm} $ as they involve the $ p-n $ loop, when the isospin symmetry is broken, the current is only partially conserved. The Fermi sea polarization function $ \Pi_{\mu\nu}^{D} $ consists of the vector-vector, vector-tensor, and tensor-tensor contributions and hence can be expressed as
\begin{equation}
\Pi_{\mu\nu}^{D}(k)=\Pi_{\mu\nu}^{vv}(k)+\Pi_{\mu\nu}^{vt}(k)+\Pi_{\mu\nu}^{tt}(k).
\end{equation}   

The real part of the temperature-density dependent polarization tensor in the region of stable collective modes for $ \rho^{+} $  is given by
\begin{eqnarray}
\Pi_{\mu\nu}^{D}(k)= & & g_{\rho NN}^{2}\left\lbrace \int\frac{d^{4}p}{(2\pi)^{4}}2\pi\delta(p^{2}-M_{N}^{*2}) \frac{\tau_
{\mu\nu}(p,p+k)}{(p+k)^{2}-M_{N}^{*2}}\left[ \theta(p_{0})n_{N}(p)+\theta(-p_{0})\overline{n}_{N}(p)\right]\right\rbrace \nonumber\\ 
& & +g_{\rho NN}^{2}\left\lbrace \int\frac{d^{4}p}{(2\pi)^{4}}2\pi\delta(p^{2}-M_{P}^{*2})\frac{\tau_{\mu\nu}(p-k,p)}{(p-k)^{2}-M_{P}^{*2}}\left[ \theta(p_{0})n_{P}(p)+\theta(-p_{0})\overline{n}_{P}(p)\right] \right\rbrace \nonumber\\ 
\end{eqnarray}
where,
\begin{equation}
\tau_{\mu\nu}(p-k,p)=\tau_{\mu\nu}^{vv}(p-k,p)+\tau_{\mu\nu}^{vt}(p-k,p)+\tau_{\mu\nu}^{tt}(p-k,p)
\end{equation}
and
\begin{eqnarray}
\tau_{\mu\nu}^{vv}(p-k,p)= & & 4\left[(p-k)_{\mu}p_{\nu}+p_{\mu}(p-k)_{\nu}-(p-k).p\: g_{\mu\nu}+M_{P}^{*2}g_{\mu\nu} 
\right],\nonumber\\
\tau_{\mu\nu}^{vt}(p-k,p)= & & 4M_{P}^{*}\frac{\kappa_{\rho}}{M}q^{2}\textsl{K}_{\mu\nu},\nonumber\\
\tau_{\mu\nu}^{tt}(p-k,p)= & & 16\left(\frac{\kappa_{\rho}}{4M_{P}^{*}} \right)^{2}\left[\textsl{K}_{\mu\nu}\left\lbrace2(p.k)^{2}-p^{2}k^{2}-k^{2}(p.k)-k^{2}M_{P}^{*2} \right\rbrace-2k^{2}\textsl{P}_{\mu\nu} 
 \right]  \nonumber\\  
\end{eqnarray}
Here $ \textsl{K}_{\mu\nu}= -g_{\mu\nu}+\frac{k_{\mu}k_{\nu}}{k^{2}}  $ and $ \textsl{P}_{\mu\nu}=\left\lbrace p_{\mu}-\left(\frac{p.k}{k^{2}} \right)k_{\mu}\right\rbrace \left\lbrace p_{\nu}-\left(\frac{p.k}{k^{2}} \right)k_{\nu}\right\rbrace   $.

It can be seen that due to the symmetry breaking, the vector current is not conserved for the $ \rho^{\pm} $. 
An evaluation of the polarization tensor shows that $ k^{\mu}\Pi_{\mu\nu}^{D}\neq 0 $ as required by the current conservation.   
By using the techniques of current algebra, it has been shown in \cite{Mazumder} that apart from the fluctuation, there arise an additional contribution to the current, which is proportional to the difference between the neutron and the proton number densities, when the ground state does not respect the $p\longleftrightarrow n $ symmetry. i.e. $ k^{\mu}\Pi_{\mu\nu}^{D}=2g_{\rho NN}^{2}(\rho_{N}-\rho_{P}) $ for $ \rho^{+} $. So a redefinition of the polarization tensor is required for studying the vector meson propagation in asymmetric nuclear matter. The modified polarization tensor in the case of the vector meson interacting with real particle-hole excitations in the nuclear medium is given by 
\begin{equation}
\tilde{\Pi}_{\mu\nu}=\Pi_{\mu\nu}\mp \frac{2g_{\rho NN}^{2}(\rho_{N}-\rho_{P})}{k_{0}}\delta_{0\mu}\delta_{0\nu} ,  
\end{equation}
for $ \rho^{\pm} $. For $ \rho^{+} $, the results are
\begin{eqnarray}
(\Pi_{\mu}^{\mu})^{D,vv}= & &\left(\frac{g_{\rho NN}^{2}}{2\pi^{2}}\right)\int_{0}^{\infty} \frac{p^{2}dp}{E_{P}^{*}(k)} \frac{1}{p|\vec{k}|}\left( n_{P} +\bar{n}_{P} \right) \left[\left( 2M_{P}^{*2}+k^{2}\right) \ln a-4p|\vec{k}|\right] \nonumber\\
& &+ \left(\frac{g_{\rho NN}^{2}}{2\pi^{2}}\right)\int_{0}^{\infty} \frac{p^{2}dp}{E_{N}^{*}(k)} \frac{1}{p|\vec{k}|}\left( n_{N} +\bar{n}_{N} \right)\left[\left( 2M_{N}^{*2}+k^{2}\right) \ln b-4p|\vec{k}|\right]  \nonumber\\
(\Pi_{\mu}^{\mu})^{D,vt}= & & -\frac{3g_{\rho NN}^{2}}{\pi^{2}} \left( \frac{k^{2}}{|\vec{k}|}\right) \frac{\kappa_{\rho}M_{P}^{*}}{2M} \int_{0}^{\infty} \frac{pdp}{E_{P}^{*}(k)}\left( n_{P} +\bar{n}_{P} \right)\ln a\nonumber\\ & &-\frac{3g_{\rho NN}^{2}}{\pi^{2}}\left( \frac{k^{2}}{|\vec{k}|}\right) \frac{\kappa_{\rho}M_{N}^{*}}{2M} \int_{0}^{\infty} \frac{pdp}{E_{N}^{*}(k)}\left( n_{N} +\bar{n}_{N} \right)\ln b  \nonumber\\
(\Pi_{\mu}^{\mu})^{D,tt}= & &\left( \frac{g_{\rho NN}^{2}}{4\pi^{2}}\right)\left(\frac{\kappa_{\rho}}{M} \right)^{2} \int_{0}^{\infty} \frac{p^{2}dp}{E_{P}^{*}(k)}\left( n_{P} +\bar{n}_{P} \right)\nonumber\\
& & \left[\frac{k^{2}}{4p|\vec{k}|}\left\lbrace \left(k^{2}+8M_{P}^{*2}\right)\ln a-4p|\vec{k}| \right\rbrace +4E_{P}^{*}k_{0}\right] \nonumber\\
& &+ \left( \frac{g_{\rho NN}^{2}}{4\pi^{2}}\right)\left(\frac{\kappa_{\rho}}{M} \right)^{2} \int_{0}^{\infty} \frac{p^{2}dp}{E_{N}^{*}(k)}\left( n_{N} +\bar{n}_{N} \right)\nonumber\\
& & \left[\frac{k^{2}}{4p|\vec{k}|}\left\lbrace \left(k^{2}+8M_{N}^{*2}\right)\ln b-4p|\vec{k}| \right\rbrace -4E_{N}^{*}k_{0}\right] 
\end{eqnarray} 

The redefined polarization tensor calculated using equation (3.15) now fullfils the condition of current conservation. $ \Pi_{L} $ and $ \Pi_{T} $ are calculated using equation (3.3). We have shown only the results for the $ \rho^{+} $ meson.  $ \rho^{-} $ being the anti-particle, the results can be obtained by replacing $ k_{0} $ with $ -k_{0} $ and $ \vert\vec{k}\vert $ with $ -\vert\vec{k}\vert $ and $ P\longrightarrow N $. From the expressions it is clear that the polarization tensors are modified differently for the different members of the multiplet and this subtle requirement when not considered would lead to unphysical splitting of the transverse and the longitudinal modes.

Here $ p = |\vec{p}| $. $ \vec{k} $ is the 3- momentum of the vector meson, $ \rho $. A and B are defined as
\begin{eqnarray}
a & = & \frac{k^{2}+2p|\vec{k}|-2E_{P}^{*}k_{0}}{k^{2}-2p|\vec{k}|-2E_{P}^{*}k_{0}}, \nonumber\\
b & = & \frac{k^{2}+2p|\vec{k}|+2E_{N}^{*}k_{0}}{k^{2}-2p|\vec{k}|+2E_{N}^{*}k_{0}}, 
\end{eqnarray} 

The real and the imaginary parts are calculated by performing the analytic continuation $ k_{0}\longrightarrow E+i\varepsilon $, where, $ E = \sqrt{M_{\rho}^{2}+|\vec{k}|^{2}} $, $ M_{\rho} $ being the invariant mass of the $ \rho $ in the medium. The spectral function is obtained as 
\begin{equation}
A_{L(T)}= - \frac{Im\Pi_{L(T)}}  {[M_{\rho}^{2}-(m_{\rho}^{2}+Re\Pi_{L(T)})]^{2}+[Im\Pi_{L(T)}]^{2}}, 
\end{equation}
where  $ \Pi_{L(T)} $ corresponds to the particular member of the triplet. Imaginary parts exist for $ M_{\rho}\geq (M_{N}^{*}+M_{P}^{*})$ \cite{PCR,Chin} for $ \rho\pm $ and $ M_{\rho} \geq 2M^{*} $ for $ \rho^{0} $.

\section{Results and Discussions}
\label{results}
The results of the calculations for the spectral function of the $ \rho $ multiplet
in the hot and dense asymmetric nuclear matter are presented in this section. 
It is clearly emphasized that the present phenomenological model is intended to 
study only the effects of the decreasing neutron and proton effective masses with neutron
and proton densities on the $ \rho $-meson spectral function in the medium. 
Also any other decay channels which can contribute to the real and the imaginary 
parts are neglected here. In the present model the effects of the finite neutron 
(proton) density enters through the nucleon loops. From the plots of the spectral 
function it will be evident that this is an important effect. The decreasing 
effective neutron (proton) mass with density and asymmetry considerably changes the $\rho$-meson 
spectral function.
 
In the chiral SU(3) model the parameters used in the calculation of the 
effective neutron (proton) masses and the corresponding effective chemical 
potentials are,
$ m_{\pi}=0.1396 $ GeV, $ m_{K}=0.498$ GeV, $ m_{\omega}=0.783$ GeV, 
$f_{\pi}=0.0933$ GeV, $f_{K}=0.122$ GeV, $\zeta_{0}=0.10656$ GeV, $k_{0}= 2.37$, 
$k_{1}= 1.4$, $k_{2}= -5.55$, $k_{3}= -2.64$, $d=0.064  $, $\chi_{0}=0.4027$ 
GeV, $g_{\sigma N}=10.6$, $g_{\zeta N}=-0.47$, $ g_{\delta N}=2.5 $, $g_{4}=2.7 $ \cite{Zsch1}.

As mentioned above, the variation of the effective neutron (proton) mass and their 
effective chemical potentials determine the spectral function of the $\rho$- meson triplet 
in the hot and dense asymmetric nuclear matter. The nuclear matter saturation density is chosen to 
be $\rho_{0}=0.15 fm^{-3}$ \cite{Zsch1}. Figure 1 depicts the variation of the effective masses of the $ \rho $ multiplet, the neutron and the proton against the scaled density for two different asymmetry parameter values and for two temperatures, $ T $=0 and 0.15 GeV within MFA. For the same temperature, there is a considerable splitting in the effective masses of the $ \rho $ multiplet, when the asymmetry parameter is increased. Among the isospin multiplet, the $ \rho^{+} $ takes a larger mass, $ \rho^{-} $ lower mass with that of $ \rho^{0} $ in between. It can be seen that the change in the effective vector meson masses with the asymmetry parameter is larger than their modification with temperature.
 
The effect of the isospin asymmetry on the 
properties of the $ \rho $ vector meson can be understood in the limit of maximum asymmetry, 
i.e. $ \eta $ = 0.5, neutron matter.

\begin{figure}[h]
\includegraphics[width=13cm]{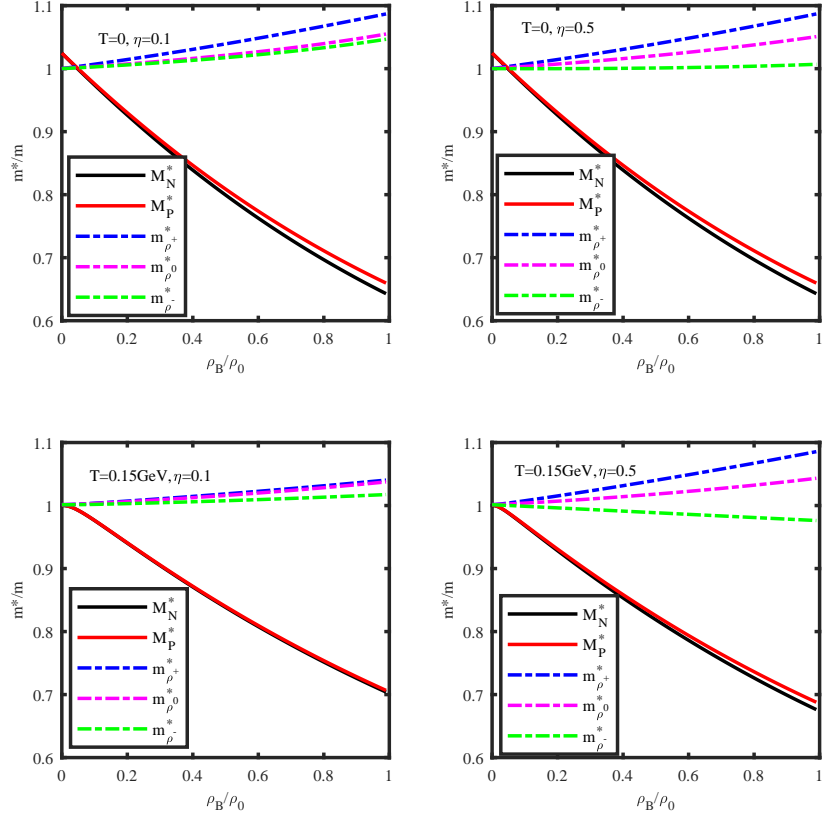} 
\caption{(Color online) Variation of  $\rho $ and nucleon effective masses against density for 
(a) $ T=0 $ and $ \eta = 0.1  $,
(b) $ T=0 $ and $ \eta = 0.5 $, (c) $ T=0.15 $ GeV and $ \eta = 0.1  $ and
(d) $ T=0.15 $ GeV and $ \eta = 0.5  $ within MFA in the chiral SU(3) model.
as functions of temperature $ T  $ in GeV.} 
\label{fig.1}
\end{figure}

\begin{figure}[h]
\includegraphics[width=13cm]{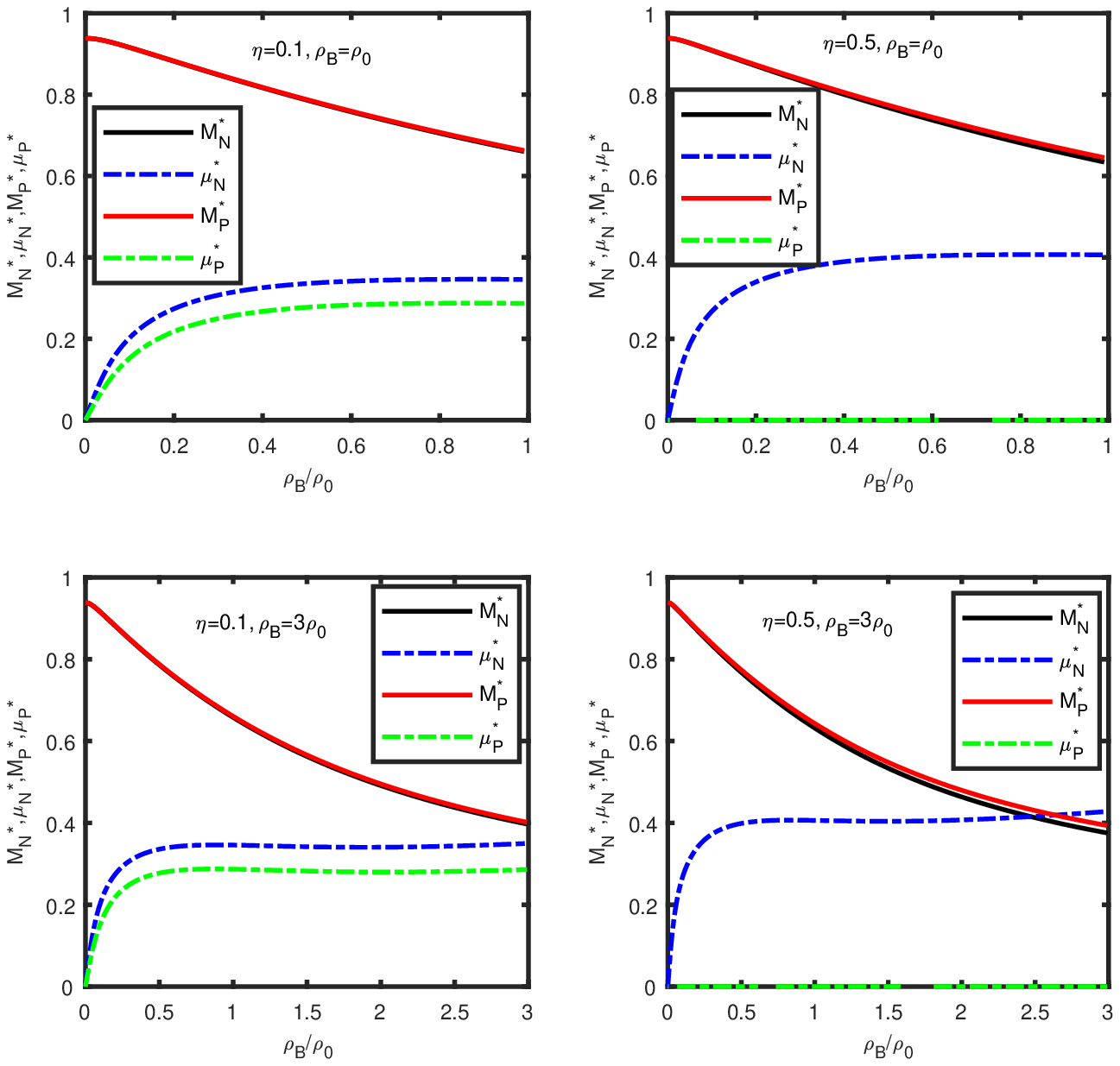} 
\caption{(Color online)Variation of nucleon effective masses and effective chemical
potentials against density for 
(a) $\rho_{B}=\rho_{0}  $ and $ \eta = 0.1  $,
(b) $\rho_{B}=\rho_{0}  $ and $ \eta = 0.5 $, (c) $\rho_{B}=3\rho_{0} $  and $ \eta = 0.1  $ and
(d) $ \rho_{B}=3\rho_{0}$  and $ \eta = 0.5  $ within MFA in the chiral SU(3) model. } 
\label{fig.2}
\end{figure}

From figure 2, it can be seen that for a particular value of the asymmetry parameter $ \eta $, 
the effective masses $ M _{i}^{*}$, (where $ i= p,n $ ) drop with the increase in density of 
the nuclear medium. This drop in the value of $ M _{i}^{*}$ with density is seen to be considerably larger 
than the modification of $ M _{i}^{*}$ due to temperature at a particular density. 
For the same density, if $ \eta $ is increased, the neutron and the proton masses 
become non-degenerate, the neutron (proton) mass is lowered (enhanced)  and this non-degeneracy 
is increased with increase in density. The neutron and the proton chemical potentials  
$ \mu_{N}^{\ast} $ and $ \mu_{P}^{\ast} $ separate, with $ \mu_{N}^{\ast} $ going 
up and $ \mu_{P}^{\ast} $ going down with increasing $ \eta $. This behaviour is 
due to the fact that at finite densities and for isospin asymmetric nuclear matter, the 
scalar-isovector field $ \delta $ contributes, whereas for isospin symmetric nuclear
matter, the $ \delta $ field has no contribution. While the condensed scalar $ \sigma $ 
and $ \zeta $ meson fields generate a shift of the nucleon mass, the $ \delta $- meson is responsible 
for the splitting of proton and neutron effective masses. The $ \delta $-meson is 
considered to be an useful degree of freedom in the description of asymmetric
nuclear matter.

The present work is limited to the discussion of the effects of the density and asymmetry 
on the spectral function of the  $ \rho $ multiplet. 
The parameters chosen in the calculation of the spectral function are, 
$ m_{\rho} = 0.77 $GeV, $g_{\rho NN}^{2}=6.96$, and $\kappa_{\rho NN}=6.1 $. 
The coupling constants $ g_{\rho NN} $ and $ \kappa_{\rho NN} $ 
are determined from the fitting to the nucleon-nucleon scattering data done by the 
Bonn group \cite{Mach}.

\begin{figure}
\includegraphics[width=13cm] {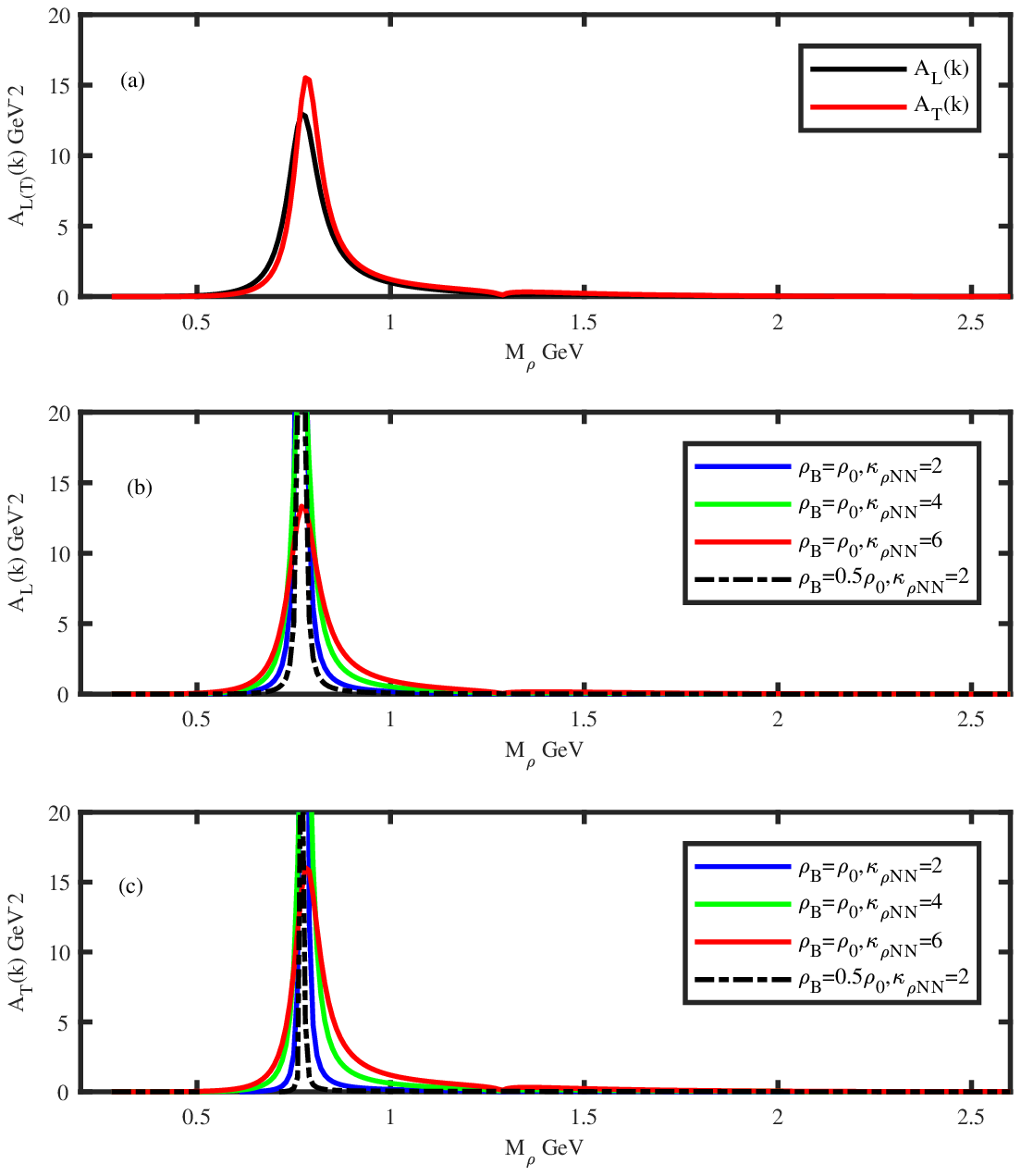} 
\caption{(Color online)Longitudinal and transverse spectral functions for the $\rho $ 
against the invariant mass $ M_{\rho} $ for  $ \rho_{B}=\rho_{0} $ in symmetric nuclear matter. 
$ | \vec{k}| $ = 0.75 GeV and $ T $=0.15 GeV.} 
\label{fig.3}
\end{figure}

\begin{figure}
\includegraphics[width=13cm]{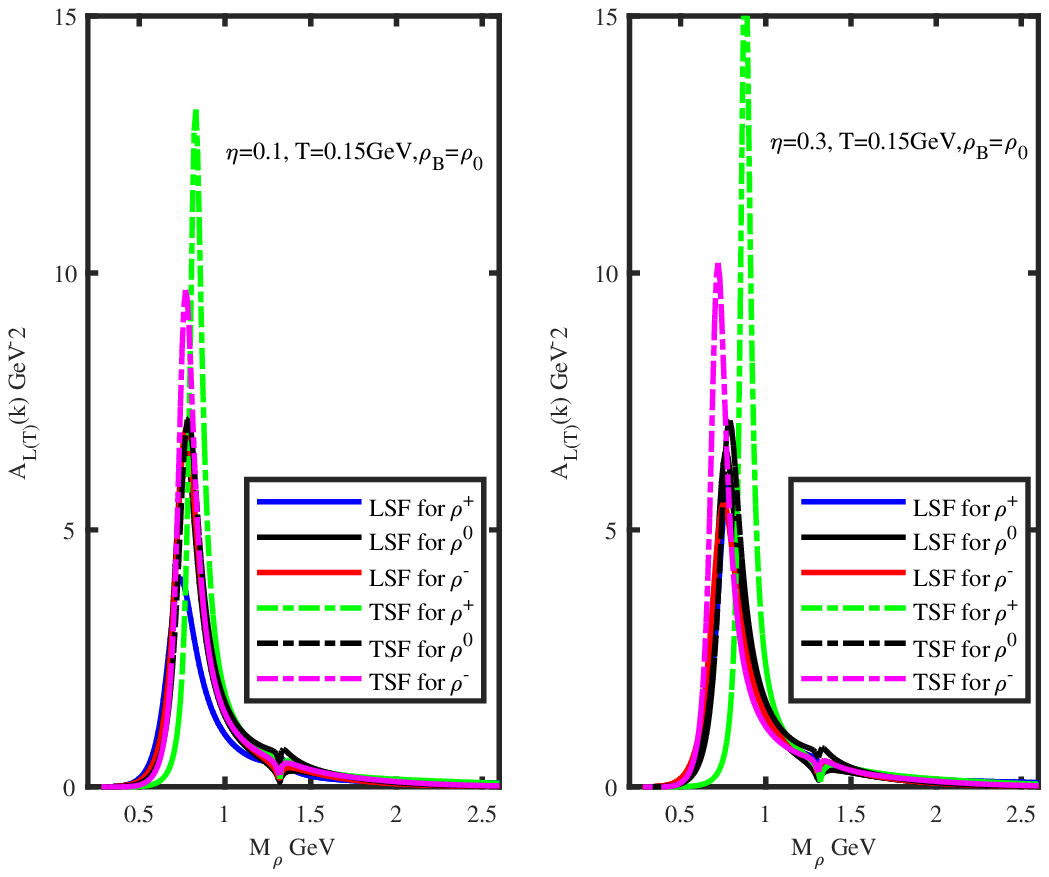} 
\caption{(Color online)Longitudinal and transverse spectral functions for the $\rho $ multiplet
against the invariant mass $ M_{\rho} $ for  $ \rho_{B}=\rho_{0} $ for two
different asymmetry parameters $ \eta $=0.1 and 0.3 respectively. $ | \vec{k}| $ = 0.75 GeV 
and $ T $=0.15 GeV.} 
\label{fig.4}
\end{figure}

\begin{figure}
\includegraphics[width=13cm]{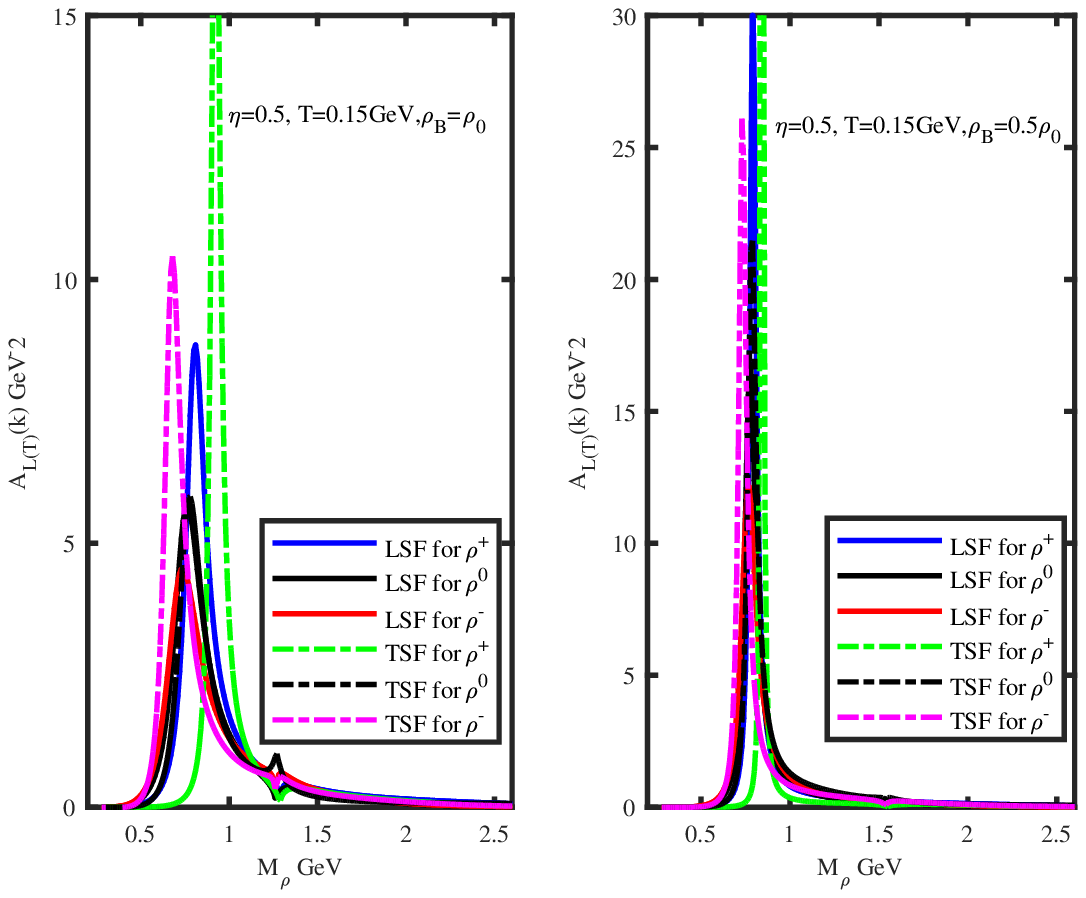} 
\caption{(Color online)Longitudinal and transverse spectral functions for the $\rho $ multiplet
against the invariant mass $ M_{\rho} $ for $ \eta $=0.5 (neutron matter) for two densities $ \rho_{B}=\rho_{0} $ and $ \rho_{B}=0.5\rho_{0} $ . 
$| \vec{k}| $ = 0.75 GeV and $ T $=0.15 GeV.} 
\label{fig.5}
\end{figure}

Figures 3, 4, and 5 plot the longitudinal and the transverse spectral functions for the $ \rho $ meson in the 
symmetric and  asymmetric cases. For $ \eta $ = 0, the longitudinal and the transverse peaks are at
0.77 GeV and 0.78 GeV respectively, at a $ \rho $ propagation momentum $ | \vec{k}| $ = 0.75 GeV for 
$ \rho_{B}=\rho_{0} $. For a small asymmetry, $ \eta $ = 0.1, in the left pannel (figure 4), there is a very small splitting between the longitudinal and the transverse spectral functions among the $ \rho $ multiplet. The longitudinal spectral function peaks for the $ \rho^{0} $ and the $ \rho^{-} $ are at 0.77 GeV, but that for $ \rho^{+} $ is at a lower value of 0.74 GeV. The transverse spectral function peaks 
separate with $ \rho^{+}$ taking a higher $ M_{\rho} $ value than $ \rho^{-} $ while that of $ \rho^{0} $ 
lies in between.      

From figures 4 and 5, one can see that for low values of $ \eta $, $ A_{L}(k)$ peaks for $ \rho^{+} $ is at a lower $ M_{\rho} $, while that for $ \rho^{0} $ and $ \rho^{-} $ are degenerate. But with an increase in $ \eta $, the peaks clearly split with that of $ \rho^{+} $ taking higher $ M_{\rho} $ values. The peak for $ \rho^{0} $ lies in between those for $ \rho^{+} $ and $ \rho^{-} $. The splitting between the transverse and the longitudinal modes of the multiplet is seen to increase with increasing values of the asymmetry. 

Now consider the extreme limit of asymmetry, i.e. the neutron matter, in figure 5.
There is a clear separation of both the modes among the isospin multiplet. 
For $ \eta $ = 0.5, $ \rho_{P} $ = 0 which implies $n_{p} = \bar{n}_{p}$ as $ \mu_{P}^{*} $ = 0, at finite T. So considering the Fermi sea contribution alone, for $ \rho^{+} $, one can see 
that the virtual proton propagator modification is due to temperature alone, whereas for the 
virtual neutron propagator, modification is due to both density and temperature. $ \rho^{+} $ 
and $ \rho^{-} $ will be modified differently as for $ \rho^{+} $ , the contribution is due to 
$ p-n_{h} $ whereas for $ \rho^{-} $, the contribution is due to $ n-p_{h} $. In the case of $ \rho^{0} $, the contributions are from $ n-n_{h} $ and $ p-p_{h} $. It can be inferred that the modification of
propagators of the members of the multiplet is different for a given asymmetry which 
results in the different contribution from different loops involved. From figures 4 and 5,
it is evident that the splitting among the different members of the multiplet and thereby
the splitting of the transverse and the longitudinal modes are highly dependent on 
the density of the nuclear matter as well as the asymmetry of the medium. 
For small asymmetry, longitudinal spectral density for $ \rho^{+} $ is wider. 
With increasing asymmetry, the contribution from the particle-hole sea ($ p-n_{h} $) becomes large and 
at maximum asymmetry the baryon density is entirely due to neutron alone. The spectral function
plot becomes narrower implying a lower decay width and more stability. 
The transverse spectral density is very narrow for low asymmetry which approaches a $ \delta $-function for maximum asymmetry implying high stability. At the extreme value of asymmetry, for $ \rho^{-} $ as mentioned above, there is no density contribution to the virtual proton propagator. 
This is reflected in the spectral function plots of $ \rho^{-} $, with $ A_{L}(k) $ getting broader and
$ A_{T}(k) $ getting narrower, implying different behaviour of $ A_{L}(k) $ and $ A_{T}(k) $.
For maximum asymmetry, the longitudinal spectral density for $ \rho^{-} $ is wider than that for 
$ \rho^{0} $ and $ \rho^{+} $ showing the instability of $ \rho^{-} $ in this mode. The transverse 
spectral density is wider for $ \rho^{0} $ than for $ \rho^{\pm} $. Moreover, the $ A_{T}(k) $ peaks 
for $ \rho^{0} $ and $ \rho^{+} $  always lie above that for $ A_{L}(k) $, while for $ \rho^{-} $, $ A_{L}(k) $ peak lies above that for $ A_{T}(k) $ except for low asymmetry values. This may be due to the different loops contributing differently for a given value of asymmetry as explained earlier.

Within MFA, as pointed in \cite{PCR}, the repulsion induced by the Fermi polarization shifts the 
spectral function to higher invariant mass values. But the imaginary part coming from the particle-hole 
sector ( $ p-n_{h} $ for $ \rho^{+} $, $ n-p_{h} $ for $ \rho^{-} $ and $ n-n_{h} $, $ p-p_{h} $
for $ \rho^{0} $ ) is found to be different for the different members of the multiplet. This 
together with the change in the real part of the spectral function is responsible for the change 
in the width and the peak positions of the different members of the multiplet.

For $ \rho_{B}=0.5\rho_{0} $, there is only a small splitting in the longitudinal spectral density 
even for $ \eta $ = 0.5. The splitting in the transverse spectral density is significant among the 
isospin multiplets, showing the different behaviour of the longitudinal and transverse modes with 
varying baryon density and asymmetry when the $ \rho $ is propagating with finite momentum in an asymmetric nuclear medium.                  

Figures 3b and 3c show the spectral density for running tensor coupling $ \kappa_{\rho NN} $.
Though the peaks always remain at 0.77 GeV and 0.78 Gev for longitudinal and transverse modes 
in the three cases, $ \sqrt{m_{\rho}^{2}+Re \Pi_{L(T)}} $ is different.
For $ \kappa_{\rho NN} $ = 2, $ Re \Pi_{L} $ is negative, whereas $ Re \Pi_{T} $ is positive. A positive value for 
$ Re \Pi_{T} $ implies a small increase in the effective mass, while a negative $ Re \Pi_{L} $, a drop in the 
effective mass. For $ \kappa_{\rho NN} $ = 2 and 4, the peaks are very sharp, but
for $ \kappa_{\rho NN} $ = 6, the peak widens implying the instability of $ \rho $ with increasing $ \kappa_{\rho NN} $
(-$ Im \Pi_{L(T)} $ increases with increasing $ \kappa_{\rho NN} $) within MFA. The transverse spectral function 
peaks are at 0.78 GeV and are less wider. For a small density $ \rho_{B}=0.5\rho_{0} $, the peaks almost
approach a $ \delta $-function for $ \kappa_{\rho NN} $ = 2. One can observe that for very small densities,
both the longitudinal and the transverse peaks coincide, while the $ \sqrt{m_{\rho}^{2}+Re \Pi_{L}} $ and
$ \sqrt{m_{\rho}^{2}+Re\Pi_{T}} $ are slightly different. As -$ Im\Pi_{T} $ is always less than -$ Im\Pi_{L} $, 
the longitudinal mode is wider than the transverse one and hence less stable.

\begin{figure}
\includegraphics[width=13cm]{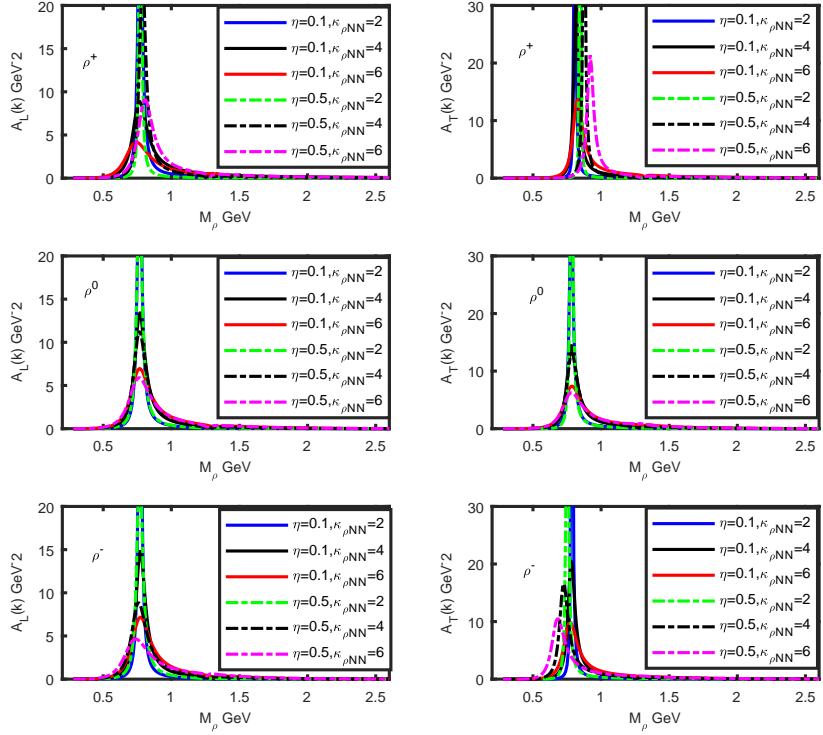} 
\caption{(Color online)Longitudinal and transverse spectral functions for the $\rho $ multiplet 
against the invariant mass $ M_{\rho} $ for $ \rho_{B}=\rho_{0} $ for two
different asymmetry parameters $ \eta $=0.1 and 0.5 respectively. $ \kappa_{\rho NN} $=2,4,6 ,  
$ | \vec{k}| $ = 0.75 GeV and $ T $=0.15 GeV.  } 
\label{fig.6}
\end{figure}

In figure 6, the longitudinal and the transverse spectral densities are presented at nuclear
matter saturation density, $ \rho_{B}=\rho_{0} $
for $ \eta $= 0.1 and 0.5. As in figure 3, $ \kappa_{\rho NN} $ is considered as a running parameter.This is 
intended to show how sensitive the properties of the vector mesons are to the magnetic interaction.
It has been shown in \cite{Mazumder}  that the effect of the magnetic interaction is to 
reduce the vector meson mass in relativistic Hartree approximation (RHA). But within MFA, the 
situation is different.    
Though the peaks for $ \rho^{0} $ are almost at the same $ M_{\rho} $ value for the symmetric and 
the asymmetric cases, those for $ \rho^{+} $ and $ \rho^{-} $ are highly displaced from the 
nominal pole mass value. From the plots for $ \rho^{+}$, for small asymmetries, $ A_{L}(k) $ peaks 
are seen to shift to low invariant mass values  with increasing $ \kappa_{\rho NN} $.
For the same $ \kappa_{\rho NN} $, if we increase the value of $ \eta $, $ A_{L}(k) $ peaks 
shift to higher invariant mass region and become less wider denoting more stability. But for 
$ \rho^{-} $ the trend is opposite, with the peaks shifting to low invariant mass region 
for maximum asymmetry. The effects are clearly seen in the $ A_{T}(k) $ peaks in the right pannel.         
For a given value of $ \kappa_{\rho NN} $, $ A_{T}(k)$ peaks  shift
to higher $ M_{\rho} $ values  for $ \rho^{+} $, whereas, they shift to lower 
$ M_{\rho} $ values for $ \rho^{-} $. $ \rho^{+} $ is always more stable in both modes
within MFA. It can also be inferred that the particles are more stable 
when $ g_{\rho NN}  >  \kappa_{\rho NN} $ within MFA. From the plots, one can observe that the 
transverse spectral density is more modified than the longitudinal mode in an 
isospin asymmetric nuclear medium. As explained earlier, this may be due to the different 
loops involved contributing differently for a given value of asymmetry. In effect, one can conclude 
that the tensor coupling splits the longitudinal and the transverse modes substantially.

\begin{figure}
\includegraphics[width=13cm]{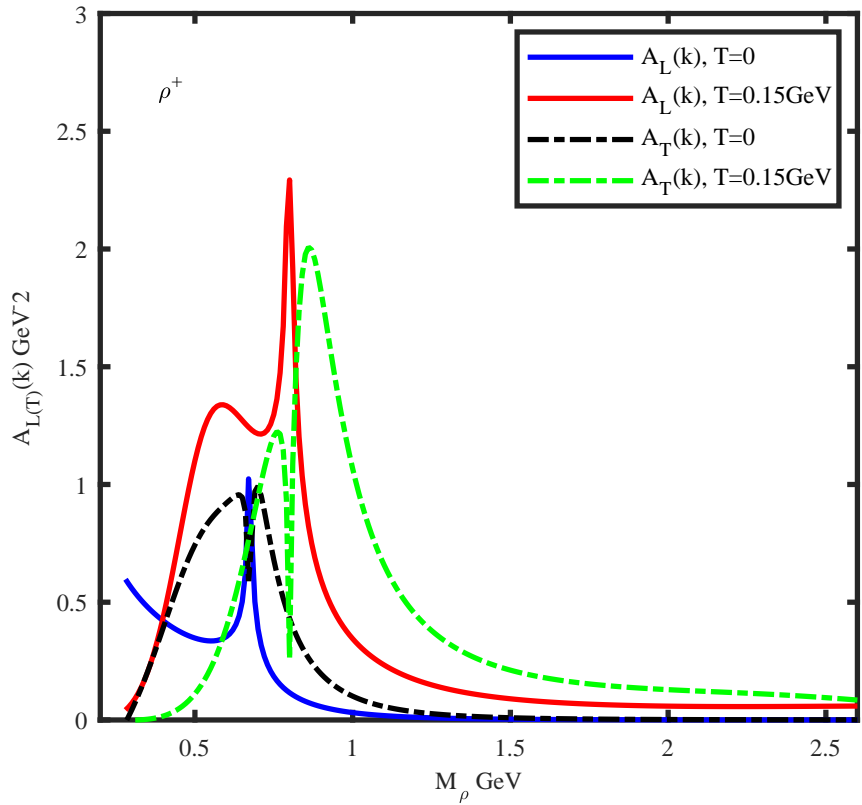} 
\caption{(Color online)Longitudinal and transverse spectral functions for $ \rho^{+} $ against the 
invariant mass $ M_{\rho} $. $ \rho_{B}=3\rho_{0} $,
$ \eta $= 0.05, $| \vec{k}| $ = 0.75 GeV, $ T $= 0 and 0.15 GeV respectively.} 
\label{fig.7}
\end{figure}

\begin{figure}
\includegraphics[width=13cm]{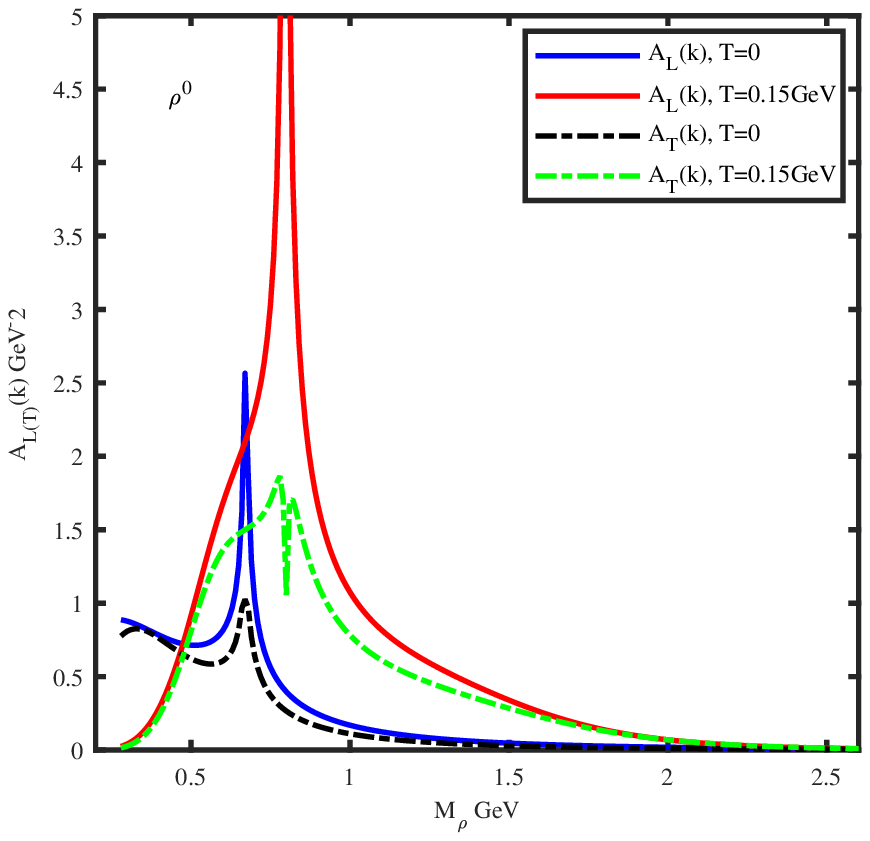} 
\caption{(Color online)Longitudinal and transverse spectral functions for $ \rho^{0} $ against the 
invariant mass $ M_{\rho} $. $ \rho_{B}=3\rho_{0} $,
$ \eta $= 0.05, $| \vec{k}| $ = 0.75 GeV, $ T $= 0 and 0.15 GeV respectively. } 
\label{fig.8}
\end{figure}

\begin{figure}
\includegraphics[width=13cm]{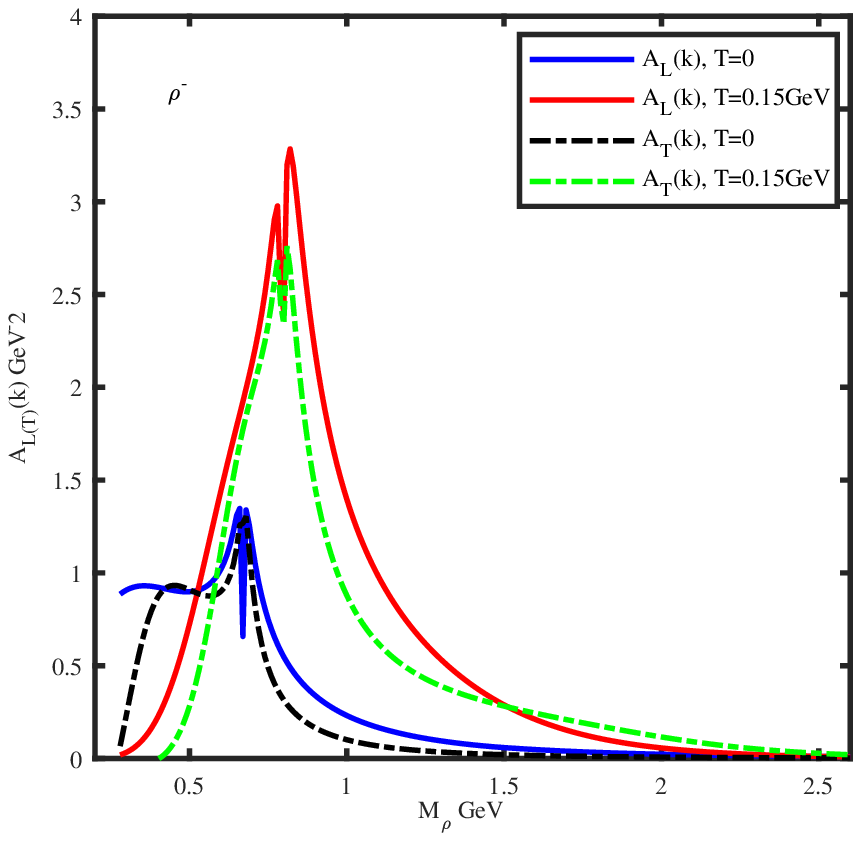} 
\caption{(Color online)Longitudinal and transverse spectral functions for $ \rho^{-} $ against the 
invariant mass $ M_{\rho} $. $ \rho_{B}=3\rho_{0} $,
$ \eta $= 0.05, $| \vec{k}| $ = 0.75 GeV, $ T $= 0 and 0.15 GeV respectively.} 
\label{fig.9}
\end{figure}

Figures 7, 8, and 9 show the spectral densities of the triplet at $ \rho_{B}=3\rho_{0} $ for     
$ T $= 0 and 0.15 GeV respectively for a very small asymmetry value $ \eta $ = 0.05 within the MFA. Even at very small asymmetries, there is a huge difference in the qualitative behaviour of the transverse and the longitudinal components of the spectral function at very high density. One can observe that
the original spectral distribution of the $ \rho $ gets broadened largely that it is no longer 
possible to interpret it as a good quasiparticle. A double-hump structure can be seen for the 
transverse spectral density for the isospin triplet at $ T $= 0.15 GeV. Such a behaviour was observed for the $ \rho $ meson in dense nuclear matter which includes baryon resonances $ N^{*}(1520) $ and $ N^{*}(1720) $ \cite{Teodorrescu}. Moreover, the broadening of the spectral density shows an accummulation of strength over a wide range of $M_{\rho}$.

\begin{figure}
\includegraphics[width=13cm]{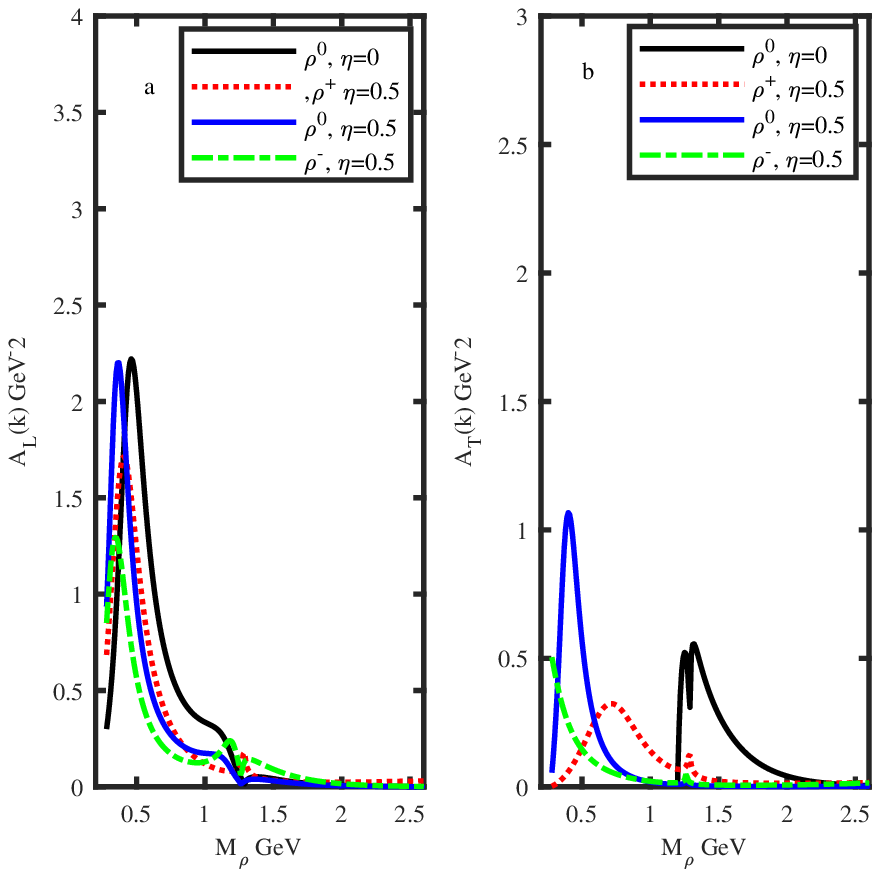} 
\caption{(Color online) Longitudinal and transverse spectral functions for the  $ \rho $ multiplet for 
nucleon matter ($ \eta $= 0 ) and neutron matter ($ \eta $= 0.5)
against the invariant mass $ M_{\rho} $. $ \rho_{B}= \rho_{0} $,
$| \vec{k}| $ = 0.1 GeV and $ T $=0.15 GeV.} 
\label{fig.10}
\end{figure}

We are studying the temperature/density effects on the spectral density when the propagation
momentum of the $ \rho $ meson is very small and the results are depicted in figure 10 for the
nuclear saturation density $ \rho_{0} $. For $ \eta $ = 0, though there is a well-defined peak 
for $ A_{L}(k)$ at $ M_{\rho} $ = 0.46 GeV, the transverse spectral function has a double-hump
like structure beyond 1 GeV. There is also a large broadening of $ A_{L}(k)$. For neutron matter ($ \eta $ = 0.5), the peaks are at 0.41 GeV, 0.37 GeV, and 0.35 GeV for $ \rho^{+} $, $ \rho^{0} $, and $ \rho^{-} $ respectively for $ A_{L}(k)$. i.e. the peaks are highly shifted towards low invariant mass region. Finite density/temperature effects are found to decrease with increasing momentum.  This may be because of the increasing many-body effects when the distances involved are larger as low momentum means larger distance \cite{Gale}. The transverse peak for $ \rho^{+} $ becomes very broad and is at $ M_{\rho} $ = 0.72 GeV. The broadening of the transverse spectral density shows an accumulation of strength towards lower invariant mass region. There is no peak for $ \rho^{-} $ in the mass range considered, while for $ \rho^{0} $ it is at 0.4 GeV. This huge difference in the qualitative behaviour of the transverse and the longitudinal modes among the members of the triplet is clearly demonstrated at low propagation momentum of the $ \rho $ meson.

\section{summary}
In summary, we have investigated the temperature and the density effects on the $ \rho $ meson
spectral function with the effective Lagrangian in the chiral SU(3) model in isospin asymmetric
nuclear medium. The spectral function of the $ \rho $ meson is studied by considering the effects
of the decreasing neutron and proton effective masses within MFA.  
The inclusion of the neutron-proton asymmetry results in the modification of the pole position 
and the imaginary parts of the $ \rho $ meson self energy in the medium, which in turn affects 
the spectral density of the particular particle concerned. Also this modification is different for 
different polarization states. 
The inclusion of the scalar-isovector field $ \delta $ which is responsible for the splitting 
of the neutron and the proton effective masses, in turn results in the splitting of the $ \rho $ meson 
triplet.   

The properties of vector mesons in nuclear matter are studied in a varied range of 
density, temperature and asymmetry. Under the effects of the Fermi sea alone, 
one can see that  the $ \rho^{+} $ takes a larger mass than $ \rho^{0} $ and $ \rho^{-} $, 
with that of $ \rho^{0} $ lying in between. At finite temperature, the masses of $ \rho^{+} $ 
and $ \rho^{0} $ undergo an enhancement, while that of $ \rho^{-} $, a reduction with 
increasing asymmetry.

The spectral densities of $ \rho $ meson propagating in asymmetric nuclear matter are 
evaluated in a mean free model. The meson spectral densities
in matter are quite different from those in free space, the difference stemming due 
to a preferred frame attached to the nuclear matter. The transverse and the 
longitudinal components of the spectral density exhibit different qualitative behaviour in 
matter, whereas they are degenerate in free space. With increasing asymmetry,
it can be seen that the splitting between the two modes of the triplet increases and 
this splitting is more pronounced for the transverse mode. While the peaks for $ \rho^{+} $ move 
to higher invariant mass regions, those for $ \rho^{-} $ move to lower invariant mass regions.
The peak for $ \rho^{0} $ is almost around its nominal pole mass value. 
Moreover, for the same asymmetry and temperature, the splitting is found to increase when the 
density is increased. A large variation in the peak positions is observed when the 
propagation momentum of the $ \rho $ is small implying that the temperature/density
effects are larger at small momentum.

The role of the magnetic interaction on the spectral properties is also analyzed by
assuming running values for $ \kappa_{\rho NN} $. The particles appear to be more 
stable when the vector coupling dominates over the tensor coupling within MFA. The 
splitting between the different modes is larger for a larger value of $ \kappa_{\rho NN} $ 
and the modification of the transverse component is larger. For large densities, within MFA
the original distribution of the $ \rho $ changes completely that it is no longer 
possible to interpret it as a particle. 

This work can be extended to include arbitrary external magnetic fields to study the variation of the  neutral and the charged $\rho$ meson effective masses, their spectral functions and the dispersion relations  with the external magnetic field.

\acknowledgements
The author gratefully acknowledges the financial support from the Department of Science \& Technology, Government of India, for the project SR/WOS-A/PS/26/2013, and Dr. Latha, Head of the Department of Physics, Providence Women's College, Calicut, for her support in carrying out this project at the Institute.

\end{document}